\documentstyle [12pt,aaspp4]{article}
\def\beq{\begin{equation}}
\def\eeq{\end{equation}}
\def\bey{\begin{eqnarray}}
\def\eey{\end{eqnarray}}
\def\pppm{\rm P^3M}
\def\mpc{\,h^{-1}{\rm {Mpc}}}

\def\msun{{M_\odot}}

\def\gs{\mathrel{\raise1.16pt\hbox{$>$}\kern-7.0pt
\lower3.06pt\hbox{{$\scriptstyle \sim$}}}}
\def\ls{\mathrel{\raise1.16pt\hbox{$<$}\kern-7.0pt
\lower3.06pt\hbox{{$\scriptstyle \sim$}}}}
\def\gtsima{$\; \buildrel > \over \sim \;$}
\def\ltsima{$\; \buildrel < \over \sim \;$}
\def\prosima{$\; \buildrel \propto \over \sim \;$}
\def\gsim{\lower.5ex\hbox{\gtsima}}
\def\lsim{\lower.5ex\hbox{\ltsima}}
\def\simgt{\lower.5ex\hbox{\gtsima}}
\def\simlt{\lower.5ex\hbox{\ltsima}}
\def\simpr{\lower.5ex\hbox{\prosima}}

\begin{document}
\title {
Accurate fitting formula for \\
the two-point correlation function of the dark matter halos}
\author {Y.P. Jing} 
\affil{Research Center for the Early Universe,
School of Science, University of Tokyo, Bunkyo-ku, Tokyo 113-0033, Japan}
\affil {e-mail: jing@utaphp2.phys.s.u-tokyo.ac.jp}
\received{---------------}
\accepted{---------------}

\begin{abstract}
 
  An accurate fitting formula is reported for the two-point
  correlation function $\xi_{hh}(r; M)$ of the dark matter halos in
  hierarchical clustering models. It is valid for the linearly
  clustering regime, and its accuracy is about 10\% in $\xi_{hh}(r;
  M)$ for the halos with mass $M>(10^{-2}\sim 10^{-3})M_{\star}$ where
  $M_{\star}$ is the characteristic non-linear mass. The result is
  found on the basis of a careful analysis for a large set of
  scale-free simulations with $256^3$ particles. The fitting formula
  has a weak explicit dependence on the index $n$ of the initial power
  spectrum, but can be equally well applied to the cold dark matter
  (CDM) cosmological models if the effective index $n_{eff}$ of the
  CDM power spectrum at the scale of the halo mass replaces the index
  $n$. The formula agrees with the analytical formula of Mo \& White
  (MW96) for massive halos with $M>M_{\star}$, but the MW96 formula
  significantly underpredicts $\xi_{hh}(r; M)$ for the less massive
  halos. The difference between the fitting and the analytical
  formulae amounts to a factor $\gs 2$ in $\xi_{hh}(r; M)$ for $M=0.01
  M_{\star}$. One of the most interesting applications of this fitting
  formula would be the clustering of galaxies since the majority of
  halos hosting galaxies satisfies $M\ll M_{\star}$.

\end{abstract}


\section {Introduction} 

It is generally believed that galaxies are formed within the deep
potential wells of the virialized dark matter (DM) halos and that
clusters of galaxies are recently collapsed objects. The study of the
physical properties of the DM halos in cosmological models therefore
provides important clues to our understanding of the Universe. In
this Letter, we report a fitting formula for the two-point correlation
function $\xi_{hh}(r)$ of the DM halos in hierarchical clustering
models. The accuracy of the fit is about $ 10\%$ in
$\xi_{hh}(r;M)$ for a wide range of halo masses.

The two-point correlation function of DM halos has been the subject of
many recent attempts at analytical modelling (e.g. Cole \& Kaiser
\cite{ck89}; Mann, Heavens, \& Peacock \cite{mhp93}; Mo \& White
\cite{mw96}, hereafter MW96; Catelan et al. \cite{clmp98}; Porciani et
al. \cite{pmlc98}) as well as of N-body simulation studies (e.g. White
et al. \cite{wfde87}; Efstathiou et al. \cite{efwd88}; Bahcall \& Cen
\cite{bc92}; Jing et al. \cite{jmbf93}; Watanabe, Matsubara, \& Suto
\cite{wms94}; Gelb \& Bertschinger \cite{gb94}; Jing, B\"orner, \&
Valdarnini \cite{jbv95}; MW96; Mo, Jing, \& White \cite{mjw96}). In
particular, using the extended Press-Schechter formalism to calculate
the correlation function of DM halos in Lagrangian space (cf. Cole \&
Kaiser \cite{ck89}) and mapping from Lagrangian space to Eulerian
space within the context of the spherical collapse model, MW96 have
derived an analytical expression for $\xi_{hh}(r;M)$:
\beq\label{linear}
\xi_{hh}(r;M)=b^2(M)\xi_{mm}(r) \,.
\eeq 
This should hold in the linearly
clustering regime where the mass two-point correlation function $
\xi_{mm}(r)$ is less than unity. The bias parameter $b(M)$ is 
\beq
\label{bias} b(M) =1+{\nu^2-1\over \delta_c}=1+{\delta_c\over
  \sigma^2(M)}-{1\over \delta_c} \,,
\eeq 
where $\sigma(M)$ is the linearly evolved rms density fluctuation of
top-hat spheres containing on average a mass $M$, $\nu\equiv
\delta_c/\sigma(M)$, and $\delta_c=1.68$ (see MW96 and references
therein for more details about these quantities). The parameter $\nu$ will be
called the peak height for convenience. Equations (1) and (2) were
found in good agreement with their N-body results by MW96 and by Mo et
al. (\cite{mjw96}), but their tests were limited to high mass halos
with $\nu \gs 1$ due to limited mass and force resolutions.
The formula has been widely used: from modeling the
correlation function of different types of galaxies (e.g., Kauffmann
et al. \cite{kns97}; Baugh et al.  \cite{bcfl98}) to interpreting the
observed clustering of various extragalactic objects (e.g., Mo et
al. \cite{mjw96}; Mo \& Fukugita \cite{mf96}; Matarrese et
al. \cite{mclm97}; Steidel et al. \cite{sadgpk98}; Coles et
al. \cite{clmm98}; Fang \& Jing \cite{fj98}).

We have measured the two-point correlation functions for the DM halos
in a large set of high-resolution N-body simulations. Each simulation
uses $256^3$ particles, and a wide range of hierarchical models are
covered: four scale-free models and three representative CDM models.
Moreover, each model is simulated with three to four different
realizations, and two different box sizes are used for each CDM model.
With these simulations of very high accuracy, we can determine
$\xi_{hh}(r,M)$ for a wide range of the halo mass $M$. As a result, we
find that the linear bias (Eq.~\ref{linear}) is
a good approximation in the linearly clustering regime, but the bias
parameter given by Eq.(\ref{bias}) agrees with the N-body results only
for massive halos with mass $M\gs M_{\star}$, where $M_{\star}$ is a
characteristic non-linear mass scale defined by
$\nu(M_{\star})=1$. For the less massive halos, the N-body results imply
significantly higher bias than the analytical prediction Eq.~(2) and the difference
in the correlation amplitude amounts to a factor $\gs 2$ for $M=0.01
M_{\star}$. Fortunately, the difference between the N-body results and
the MW96 formula can be modeled by a simple fitting formula (\S
3). This formula can fit the simulation bias parameter for halo mass
$M/M_{\star}\gs 0.01$ with an accuracy of about 5\%. 
The new findings have profound implications for the
formation of the large scale structures. One of the most interesting
applications of the fitting formula would be the clustering of
galaxies, since the local late type and dwarf galaxies are
believed to have mass $\sim 10^{11} \msun$ and to form recently
(redshift $z\ls 1$; Mo et al.\cite{mmw98}) while $M_{\star}
\sim 10^{13}\msun$ is expected for the present Universe (cf. \S3).

The simulations will be described in section 2, with emphasis on the
aspects relevant to this Letter. In section 3, we will present the
correlation function of the halos and the fitting formula. The
implications for theories and observations are discussed in section 4.

\section{Models and Simulations}

The two-point correlation functions of halos are studied both for
scale-free models and for representative CDM models of hierarchical
clustering. In the scale-free models, a power-law power spectrum
$P(k)\propto k^n$ is used for the initial density fluctuation and the
universe is assumed to be Einstein--de Sitter, $\Omega=1$. Four models
with $n=-0.5$, $-1.0$, $-1.5$, and $-2.0$ are studied. Because these
models are conceptually simple and exhibit interesting scaling
properties, it is relatively easy to understand how physical
properties depend on the shape of the power spectrum and, perhaps more
importantly, to distinguish the physical effects from numerical
artifacts, since the latter should not in general obey the scaling
relations. For this reason, we will extensively use the scaling
property that the bias parameter $b$ depends only on the halo mass $M$
scaled by $M_{\star}$, i.e. $M/M_{\star}$ for each $n$. This scaling
property manifests itself when the bias parameter for $M/M_{\star}$ is
plotted for different evolution times.  Each of our simulations for
$n\ge -1.5$ is evolved for 1000 time steps with a total of seven
outputs, and that for $n=-2.0$ is evolved for 1362 steps with eight
outputs. The output time interval is chosen so that $M_{\star}$ at
each successive output is increased by a factor 2.5, and the
$M_{\star}$ values (in units of the particle mass) at the first output
are 74 , 59, 35, and 13 for $n=-0.5$, $-1.0$, $-1.5$, and $-2.0$
respectively. Note that fixing the $M_{\star}$ values is equivalent to
fixing the normalization for the power spectra.  The time step and the
integration variables are taken similarly to Efstathiou et al.
(\cite{efwd88}). In this Letter we will rely on these scale-free
models to understand how the halo-halo correlation depends
on the shape of the power spectrum. Then we will examine whether the
result obtained from the scale-free models can be applied to CDM
models, since CDM models, at least variants thereof, are believed to
be close to reality.

Three CDM models are very typical: one is the (ever) standard CDM
model (SCDM), one is an open model with $\Omega_0=0.3$ and with a
vanishing cosmological constant $\lambda_0$ (OCDM), and the other is
a flat lower density model with $\Omega_0=0.3$ and $\lambda_0=0.7$
(LCDM). These CDM models are completely fixed with regard to the DM
clustering if the initial density power spectrum is fixed. For our
simulations, the linear CDM power spectrum of Bardeen et al.
(\cite{bbks86}) for the primordial Harrison-Zel'dovich spectrum is
used for the initial condition, which is fixed by the shape parameter
$\Gamma=\Omega_0 h$ and the amplitude $\sigma_8$ (the rms top-hat
density fluctuation on radius $8\mpc$). The values of ($\Gamma$,
$\sigma_8$) are (0.5, 0.62) for SCDM, (0.25, 1) for OCDM, and (0.20,
1) for LCDM.

Each model is simulated with our vectorized $\pppm$ code on the Fujitsu
VPP300/16R supercomputer at the National Astronomical Observatory of
Japan. Each simulation is performed with $256^3$ ($\approx 17$
million) particles and with good force resolution $\eta\approx 1/2000
L$ (where $L$ is the simulation box size). To properly understand the
effect of the cosmic variance, three to four independent realizations
are generated for each simulation of one model. Furthermore, two
different box sizes, $100\mpc$ and $300\mpc$, are adopted for each CDM
model. 
Further details about
the code and the simulations will be given in a forthcoming paper
(Jing, in preparation), where many clustering statistics of the dark
matter will also be presented. The CDM simulations with box size $100\mpc$
were used by Jing \& Suto (\cite{js98}) to study the constraints on
cosmological models of the high concentration of the Lyman Break
galaxies at redshift $z\approx 3$ discovered by Steidel et al. (1998).

\section{The correlation function of the halos and the bias parameter}
 
The DM halos are identified with the Friends-of-Friends (FOF) algorithm
with a linking parameter $0.2$ times the mean particle separation. The
halos with at least 20 members are used for the clustering
analysis. It is known that the mass defined by the members of
such FOF groups is very close to that defined by the spherical
overdensity (SO) virialization (Cole \& Lacey \cite{cl96}) and that
the mass function of such FOF groups follows the predictions of the PS
formalism (e.g. Lacey \& Cole \cite{lc94}; Mo et al. \cite{mjw96}).
More importantly, the correlation function of DM halos is quite robust
to reasonable halo identification methods, since, for example, the
correlation function of the FOF groups is statistically
indistinguishable from that of the SO groups (Mo et al. \cite{mjw96}).
Therefore, it would suffice to use the FOF groups for the present purpose.

\begin{figure} \epsscale{1.0} \plotone{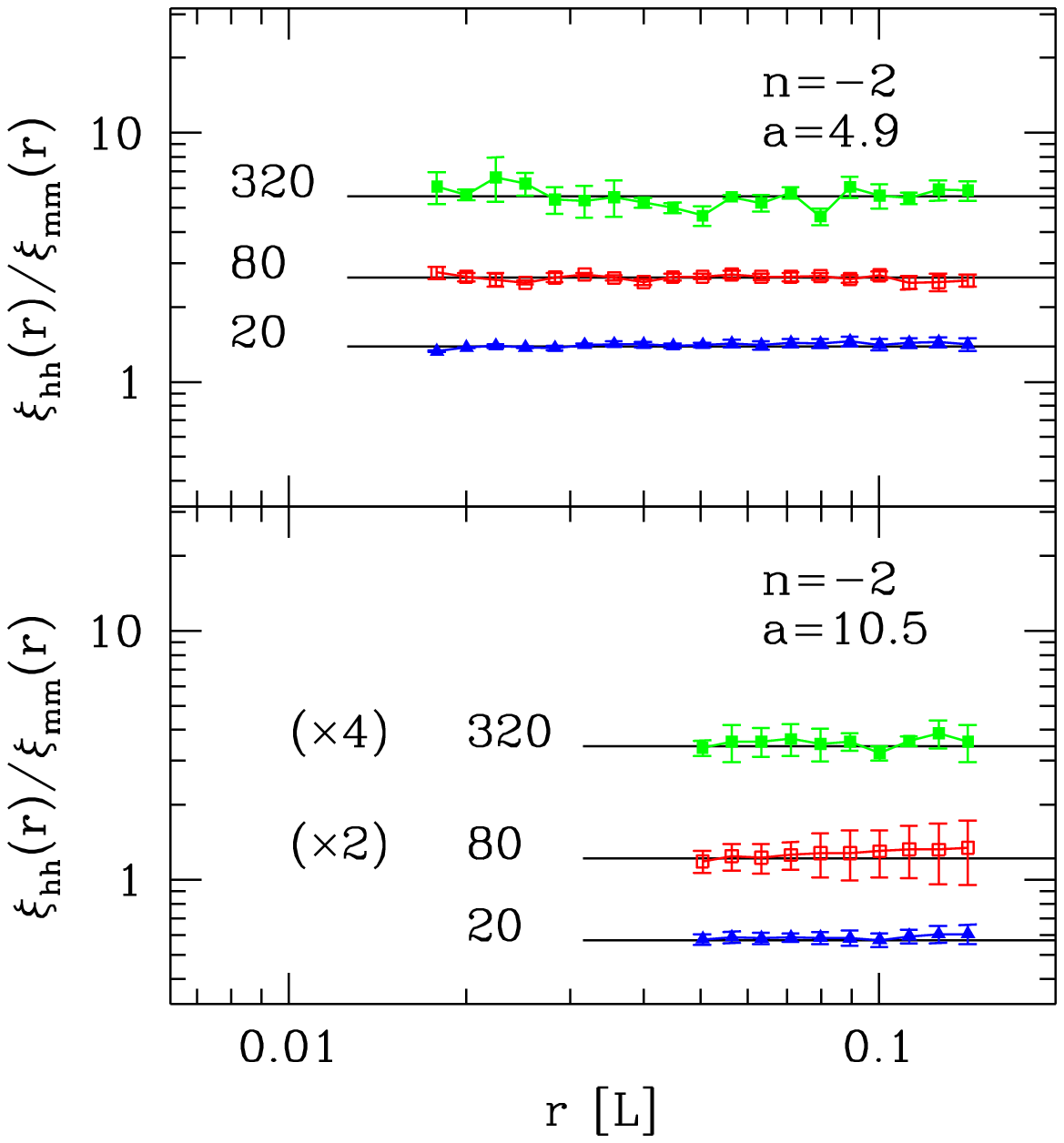} 
\caption{ The ratio as
a function of the separation $r$ of the two-point correlation function
of halos to that of the dark matter at two time outputs of the
scale-free model with $n=-2$.  The separation is in units of the
simulation box size $L$, and only the linear clustering regime is
considered. The mass of the halos $M$, in units of the particle mass,
is in the ranges $20\le M<40$ (blue triangles), $80\le M<160$ (red
open squares), and $320\le M<640$ (green filled squares) respectively. Because
the ratios of the three different masses at the later output (the
lower panel) are very close, for clarity, the ratios for the two
larger masses have been multiplied by the factors indicated in
parenthesis. The solid lines are the mean ratio averaged for different
scales.}
\label{fig1}\end{figure}

The first concern of this work is whether the bias of the halos is
linear (Eq.~\ref{linear}) in the linearly clustering regime.  We have
calculated the ratio of $\xi_{hh}(r)$ to $\xi_{mm}(r)$ for every
simulation output. Some examples which are also typical are shown in
Figure~\ref{fig1}, which show the ratio at two different outputs of
the $n=-2$ simulation for three different halo masses.  The results
are plotted only for the linear clustering regime
i.e. $\xi_{mm}(r)<1$, as only this regime is considered here.  Error
bars are calculated by averaging over the different
realizations. It is remarkable that the ratio is a constant within the
$1\sigma$ error bars, i.e. the bias is linear and the bias parameter
$b$ is a function of $M$ only. This statement is consistent with many
previous studies, but is shown here with higher accuracy.

Since $b(M)$ depends only on $M$, we predict that for a scale-free
model, if the mass $M$ is scaled by the characteristic mass
$M_{\star}$, the bias parameter $b$ depends only on the scaled mass
$M/M_{\star}$.  Interestingly, by definitions of $\nu$ and $M_{\star}$,
the scaled mass obeys a simple relation
to the peak height $\nu$, i.e.  $\nu=(M/M_{\star})^{n+3/6}$. Thus it
is also very convenient to compare the simulation data of
$b(M/M_{\star})$ with the formula of MW96 (cf. Eq.~\ref{bias}). In
Figure~\ref{fig2}, we plot the square of the bias parameter $b^2$ as a
function of the scaled mass $M/M_{\star}$. The results for all
simulation outputs are depicted. This is equivalent to a factor $\sim
10^4$ in the halo mass resolution. It is very remarkable that the
$b^2(M/M_{\star})$ results agree very well for different evolutionary
outputs.  The precise scaling behaviour given by the simulation data
assures that any numerical artifacts have negligible effect on the
results of Fig.~\ref{fig2}.

Now we compare our simulation results with the prediction of MW96. The
MW96 predictions are drawn on Fig.~(\ref{fig2}) in dotted lines.  It
is interesting to see that the MW96 formula agrees well (within
$\sim 1\sigma$ uncertainty) with the simulation results for massive halos,
i.e. $M/M_{\star}>1$. However, at $M/M_{\star}\approx 1$ the
simulation data start to deviate from the MW96 prediction, and the
deviation increases with the decrease of the scaled mass (or
equivalently with the decrease of the peak height). At
$M/M_{\star}=0.01$, the simulation result is about 2 to 4 times higher
than the prediction.  Considering the important role played by the
correlation functions of the halos in the cosmological studies, we have
searched for a fitting formula for $b(M/M_{\star})$. The formula
\beq\label{fitting}
b(M)=({0.5\over \nu^4}+1)^{(0.06-0.02n)} (1+{\nu^2-1\over \delta_c});
\hskip1cm \nu=(M/M_{\star})^{n+3/6}
\eeq
can fit the simulation results (Fig.~\ref{fig2}) with an accuracy of
about 5\% for the halo mass that the simulations can probe,
i.e. $M/M_{\star}\gs 10^{-3}$ for $n=-0.5$; $M/M_{\star}\gs 2\times
10^{-3}$ for $n=-1.0$; $M/M_{\star}\gs 3\times 10^{-3}$ for $n=-1.5$;
$M/M_{\star}\gs 10^{-2}$ for $n=-2.0$. The fitting formula recovers
the MW96 analytical formula (Eq.~\ref{bias}) at high mass
$\nu\gs 1$. The deviation of the MW96 prediction from the simulation
results at small mass is accounted for by the factor $({0.5\over
  \nu^4}+1)^{(0.06-0.02n)}$.

Since the scale-free models are only approximations to the real
Universe at some specific scales, it is very important to consider
more realistic models. In Figure~\ref{fig3}, we present the $b^2(M)$
for the halos in the three CDM models. The blue squares are
determined from the simulations of box size $300\mpc$ and the red
triangles are from those of the smaller boxes. In consistency with the
scale-free models, the simulation data are significantly higher than
the MW96 prediction (the dotted lines) for masses less than
$M_{\star}$, where $M_{\star}$ is about $10^{13} h^{-1}\msun$ for the SCDM
and about $2\times 10^{13} h^{-1}\msun$ for the low density models. Since the
slopes of the CDM power spectra change with scale, the fitting
formula~(\ref{fitting}) is not directly applicable.  Fortunately, the
fitting formula depends very weakly on the power index $n$, and it can
describe the CDM data very accurately if an effective index $n_{eff}$
replaces $n$ in Eq.~(\ref{fitting}) and the original definition of 
$\nu=\delta_c/\sigma(M)$ is used for $\nu$.  The effective index is
defined as the slope of $P(k)$ at the halo mass $M$
\beq \label{neff}
n_{eff}={d \ln P(k) \over d \ln k}\Bigg|_{k={2\pi\over R}}; \hskip 1.5cm
R=\Bigl({3M\over 4\pi\bar\rho}\Bigr)^{1/3}\,, 
\eeq 
where $\bar\rho$ is the mean density of the universe. The solid lines
in Fig.~(\ref{fig3}) are predicted in this way. They agree very well
with the simulation results. The fact that $n_{eff}$ changes very
slowly with the mass M for $M\le M_{\star}$ might be the main reason
why the fitting formula can work very well for CDM models after the simple
modification.

\section{Discussion and conclusions}
\begin{figure}
\epsscale{1.0} \plotone{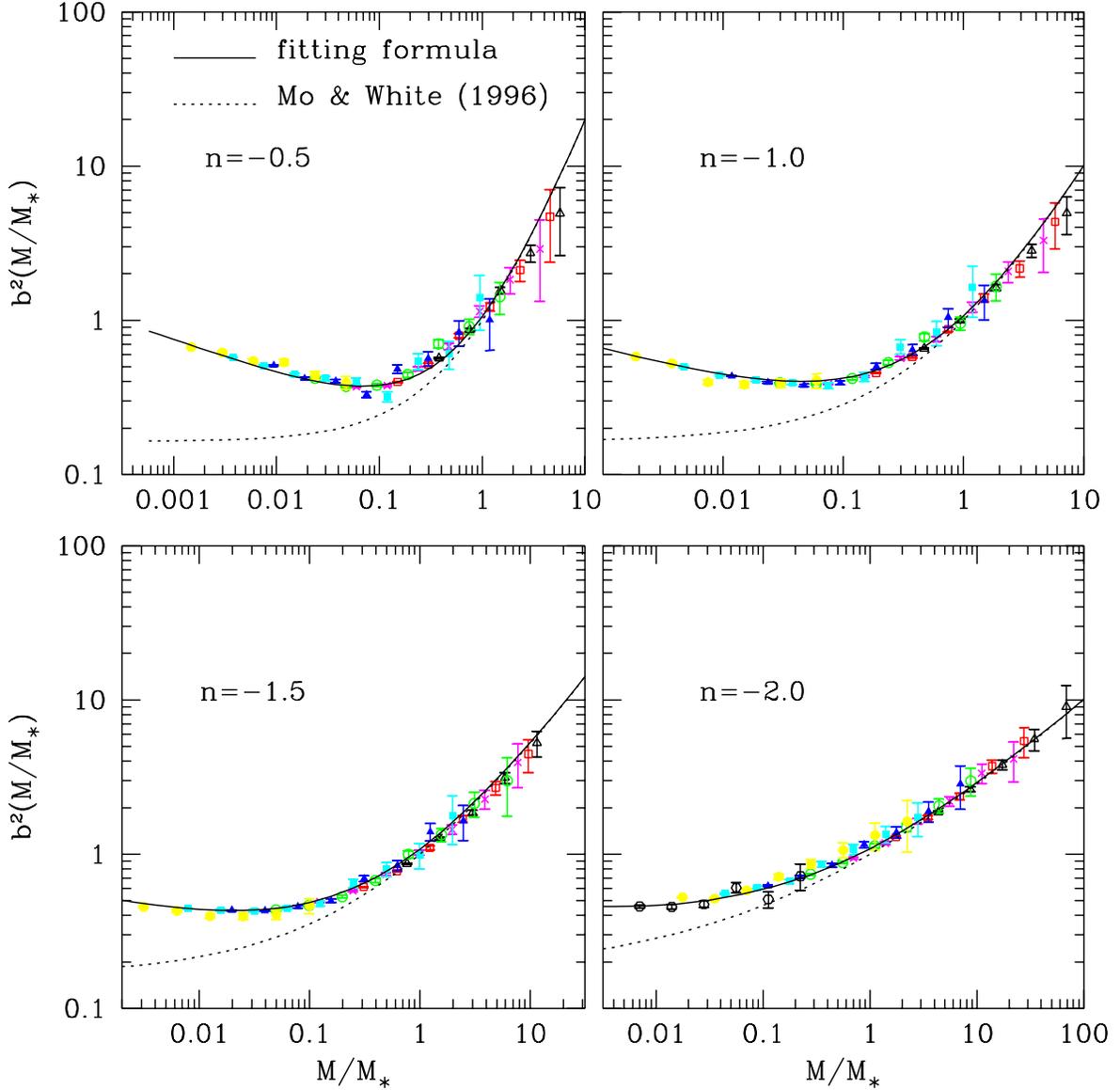}
\caption{ The square of the bias parameter as a function of the scaled
mass. The results at seven different evolutionary stages are plotted
with differently colored symbols. From the early to the late outputs,
the symbols are respectively black open triangles, red open squares,
magenta crosses, green open circles, blue solid triangles, cyan solid
squares, and yellow solid circles. For $n=-2$, the result for a
further output (more clustered) at the 1362th time step is added with
black hexagons. It is interesting to note that the results from
different outputs agree remarkably well. The dotted lines are the
prediction of Mo \& White (\cite{mw96}), which is in good agreement
with the simulation results for $M/M_{\star}\gs 1$ while it
significantly underpredicts for $M/M_{\star}\ll 1$. The solid lines
are from the simple formula found in this paper (Eq.~\ref{fitting}) which can
accurately fit the simulation results.  }\label{fig2}\end{figure}

\begin{figure} \epsscale{1.0} \plotone{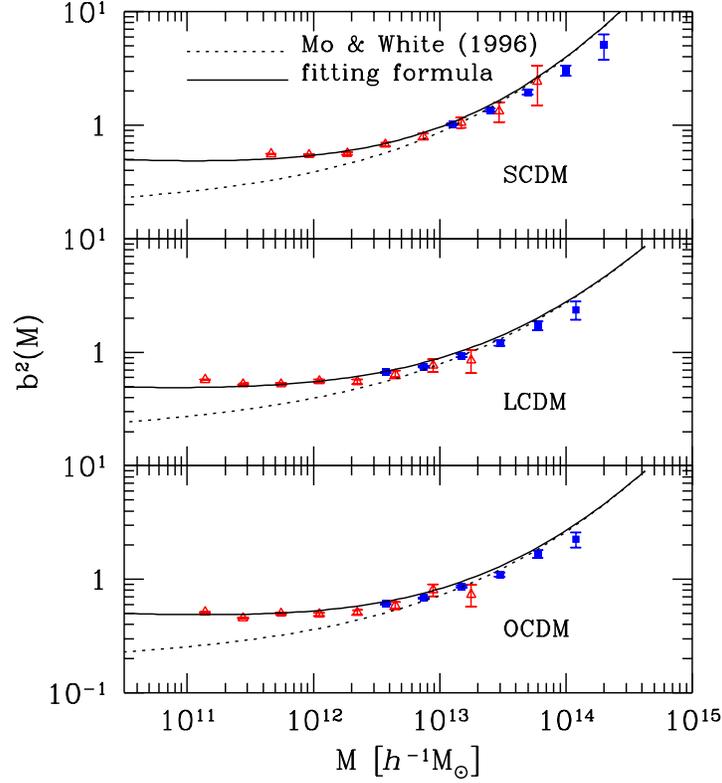} \caption{ The square
of the bias parameter as a function of the halo mass for three typical
CDM models. The blue squares are for the simulations of the larger
box, and the red triangles are for the smaller box. The dotted lines
are the prediction of Mo \& White (\cite{mw96}), which showed a
similar behavior to that found in the scale-free models when compared
with the simulation data. The solid lines, which agree quite well with
the simulation data, are the prediction of the fitting formula in this
paper with the index $n$ in Eq.(\ref{fitting}) replaced with the
effective one $n_{eff}$ at the halo mass scale (see text).  }
\label{fig3}\end{figure}

We have determined the correlation function of the DM halos for the
hitherto largest set of high-resolution N-body
simulations. The excellent scaling exhibited in the bias parameter
$b(M/M_{\star})$ at the different evolutionary outputs for the
scale-free models assures that the results in Figures~\ref{fig2} and
\ref{fig3} are physical, not contaminated by numerical
artifacts. The simulation results are in good agreement with the
formula of Mo \& White (1996) for massive halos $M\gs M_{\star}$.
However, for less massive halos the simulation results are
significantly higher. The MW96 formula was found in good agreement
with the results of $100^3$-particle scale-free simulations by MW96 and
with the results of $128^3$-particle CDM simulations of box size $\sim
300\mpc$ by Mo et al.(1996). However, their tests were limited to
halos with $M\gs M_{\star}$ because of the relatively poorer mass
resolutions. The results found here therefore do not contradict with,
but in fact support and extend the previous N-body tests.

The fitting formula (\ref{fitting}) we found for $b(M)$ is accurate
for halo masses $M/M_{\star}> 10^{-2}\sim 10^{-3}$ with only about 5\%
error. The formula can be applied both to the scale-free models and to
the CDM models. In the latter case, the index $n$ in equation
(\ref{fitting}) should be replaced with the effective power spectrum
index $n_{eff}$ (Eq.~\ref{neff}) and the original definition of
$\nu=\delta_c/\sigma(M)$ is used for $\nu$ . This fitting formula
could have many important applications for the studies of the
large-scale structures. One of them would be to predict and to
interpret the clustering of late type galaxies and dwarf galaxies both
in real observations and in analytical modeling of galaxy formation,
since these galaxies are believed to have formed recently ($z\ls 1$;
cf Mo et al. \cite{mmw98}) with halo masses much less than
$M_{\star}$.

At present we do not know the {\it exact} reasons which cause the MW96
formula to fail at the small halo masses $M/M_{\star}\ll 1$. In the
derivation of MW96, two main assumptions are 1) the halo formation is
determined by the peak height through the extended Press-Schechter
formalism; 2) the mapping of the halo clustering pattern from the
Lagrangian space to the Eulerian one is local and linear (with the
spherical collapse model). The first assumption could break down more
seriously for low peak-height halos, because the tidal force plays a
more important role in their formation. Though, as Katz, Quinn \& Gelb
(\cite{kqg93}; see also Katz et al. \cite{kqbg94}) pointed out, the
peak-height is not the sole parameter even for the formation of
high-peak halos. The validity of the local linear mapping was recently
questioned by Catelan et al.(\cite{cmp98}) in a different
context. Unfortunately their result is not directly applicable to this
Letter.  In relation to the gentle rise observed for the bias
parameter at the small $M/M_{\star}$ (Fig.~\ref{fig2}), we have
visually inspected the spatial distribution for halos with
$M/M_{\star}\approx 3\times 10^{-3}$ ($b\approx 0.8$),
$M/M_{\star}\approx 10^{-1}$ ($b\approx 0.6$), and $M/M_{\star}\approx
1 $ ($b\approx 1$) in one late output of the $n=-0.5$ model. The small
and large halos appear to delineate filamentary structures more
closely than the median-mass halos, consistent with the measured
$b$. This however might hint that either or
both of the two assumptions are violated for the small halos, since
the small halos are otherwise expected to be more
preferentially located in low density regions. Obviously it would be very
interesting to find out why the MW96 formula fails. In a future paper,
we will examine this question more closely.

\acknowledgments 

I am indebted to Gerhard B\"orner and Yasushi Suto for their critical
reading of earlier versions of the manuscript. Constructive comments
by them, by Houjun Mo, and by an anonymous referee improve the
presentation of this Letter. Yasushi Suto is thanked for his
invaluable assistance and continuing encouragement; it would be
impossible to get these very large, high quality simulations without
his help.  It is also my pleasure to acknowledge the JSPS foundation
for a postdoctoral fellowship.  The simulations were carried out on
VPP/16R and VX/4R at the Astronomical Data Analysis Center of the
National Astronomical Observatory, Japan.

 \end{document}